\input psfig.sty

\documentstyle{l-aa} 
\begin{document} 
\thesaurus{13.07.1} 
 
\title{HST Data Suggest Proper Motion for the Optical 
Counterpart 
of GRB 970228. $^{\spadesuit}$}

\author{P. A. Caraveo\inst{1}, R. P. Mignani\inst{2}, M. 
Tavani\inst{1,3} and G. F. Bignami\inst{4,1}}

\offprints{P. Caraveo, pat@ifctr.mi.cnr.it} 

\institute{$^{1}$ Istituto di Fisica Cosmica del CNR, Milano, Italy \\ 
           $^{2}$ Max-Plack-Institute f\"ur Extraterrestrische Physik, Garching, Germany \\
           $^{3}$ Columbia University, New York, USA \\
           $^{4}$ Agenzia Spaziale Italiana, Roma, Italy
}   

\date{Received ....; Accepted .... } 
\maketitle 
\markboth{P.A. Caraveo et al.: HST Observations Suggest Proper Motion for GRB 970228}
\noindent
$\spadesuit$ Based on observations with the NASA/ESA Hubble Space Telescope, 
%obtained at the Space Telescope Science Institute, which is operated by 
%AURA, Inc., under NASA contract NAS 5-26555. 
%
%\vspace{0.5cm}

\begin{abstract} 
After a quarter of a century of $\gamma$-ray burst (GRB) astronomy, the Italian-Dutch satellite 
BeppoSAX on Feb $28^{th}$, 1997 detected a soft X-ray afterglow from GRB 970228 and positioned 
it accurately. 
This made possible the successful detection of an optical transient. \\ Two public Hubble Space 
Telescope 
(HST) images of the GRB/optical transient region were taken on March $26^{th}$ and April 
$7^{th}$, 1997. 
They are analyzed here, with the purpose of understanding the nature of GRB 970228.  We find 
that the position of the faint 
point-like  object ($m_{v} \sim 26$) 
seen at the transient location  changed by $0.40 \pm 0.10$ pixels in 12 
days, corresponding to a proper motion of $\sim$ 550 mas/year. 
By comparison, four adjacent sources in the same field do not show any significant displacement, 
with astrometric residuals close 
to zero and average absolute displacements less than 0.09 pixels. If confirmed, this result 
would strongly support the 
galactic nature of GRB 970228.
\end{abstract} 
\keywords{Gamma Ray Burst, optical, GRB970228} 
 
\section{Introduction.} 
GRB 970228 stands out in the history of $\gamma$ ray bursts (GRBs). For the first time, an X-ray 
afterglow from a GRB was detected and positioned accurately thanks to the Italian-Dutch satellite BeppoSAX 
(Costa et al, 1997a,b,c).  \\
This triggered a number of optical searches in the error box. 
In particular, an $m_{V}=21.3$ object found  inside the error box $\sim$ 
20 hours after the event (van Paradijs et al, 1997; Guarnieri et al, 1997) could not be detected 
($m_{V} \ge 23.6$) $\sim$ 8 days later. van Paradijs et al. (1997) proposed the optical transient to be 
the counterpart of GRB 970228. Subsequent deeper images showed a faint extended object ($m_{R} \sim 
24$) at the 
transient's location (Groot et al, 1997; Metzger et al, 1997a). If this is interpreted as a 
galaxy (van Paradijs et al, 1997), 
the "cosmological distance scale" for GRBs would be strongly favored. 
%A galaxy detectable by the New Technology Telescope and the Nordic Optical 
%Telescope (van Paradijs et al, 1997) should 
%clearly show up in  HST Wide Field and Planetary Camera 2 (WFPC2) images.   
However, deep HST images of the field (Sahu et al, 1997a,c), 
taken on March $26^{th}$ with the WFPC2, fail to clearly show the existence of a galaxy. At 
best, 
an extended emission surrounding an off-centered point-like source is visible in the V-band 
image,  while 
the I-band image is inconclusive, also owing to significantly
lower signal-to-noise. 
Indeed, it would now appear that, according to ground-based measurements taken on April 
$2^{nd}$, the nebulosity 
has faded away (Metzger et al, 1997b). This might be taken for evidence of time variability of 
the extended emission, which 
would then not be cosmological at all. Rather, it would be galactic, and the extended optical 
emission might 
be tracing the remnant of the burst event. The WFPC2 point-like source would also be nearby. It 
could, for 
example, be a local neutron star, moving at typical pulsar speeds (Lyne \& Lorimer, 1994), or a 
stellar object of different 
nature. A proper motion search is then justified since, as shown by Caraveo et al (1996) in the 
case of the 
parallactic displacement of the Geminga neutron star (d $\sim$ 160 pc and $m_{V}=25.5$), the HST 
Planetary Camera 
can measure relative displacements to better than one tenth of a pixel even    for faint 
objects. 

\section{Data reduction.}

\begin{figure} 
\centerline{\hbox{
\psfig{figure=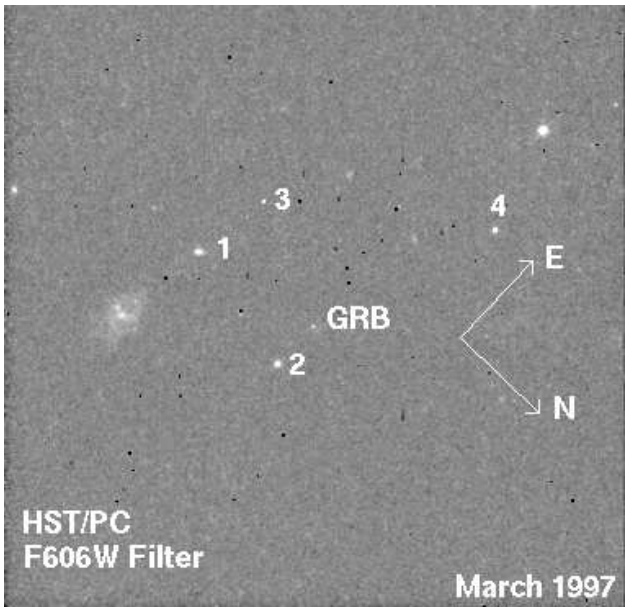,height=8cm,clip=} 
}}
%\picplace{8.cm}
\caption{HST/WFPC2 4700 sec exposure of the field of GRB 970228 using filter F606W taken on 
March $26^{th}$, 1997. 
Numbers 1 to 4 identify the field stars used for relative astrometry. The faint point source at 
the 
center of the field, labelled as GRB, is the proposed counterpart of GRB 970228.} 
 \label{} 
 \end{figure}

A second set of HST images was taken on April $7^{th}$, 1997 between 3:42-7:45 
UT, and made public after few hours (Sahu et al, 1997b).   To achieve the highest angular 
resolution, the Planetary Camera (Field of View 35" x 35", pixel size 0.0455 arcsec)
was centered on the coordinates of the optical transient. The observing strategy was identical 
to that 
used for the March observation, i.e. four F606W (an extended V filter centered at $\lambda=5843 
\AA$ with $\Delta 
\lambda =1578 \AA$) images for a total exposure time of $\sim$ 4700 s, and two F814W (roughly 
equivalent to I, centered at $\lambda = 
8269 \AA$ with $\Delta \lambda=1758 \AA$) images for a total time of 
$\sim$ 2400 s.  First, the March and April data sets have been cleaned for cosmic ray hits, 
using the IRAF/STSDAS task {\it combine}. Next, to
overcome the limitation on star centering accuracy induced by the PC undersampling of the Point 
Spread Funtion, a 3x3 box smoothing is applied.
Although based on a simple algorithm, this procedure is very effective in restoring the 
astrometric potential of the PC data (see Caraveo et al, 1996).  Figure 1 shows the March 26 
V-band field image.
The point-like 
source seen on March $26^{th}$ is well detected on April $7^{th}$. Our measured V-band magnitude 
is $26.1 \pm 0.2$ 
($26.0 \pm 0.3$, according to Sahu et al, 1997b), as opposed to the value 
of $25.7 \pm 0.3$ measured on March $26^{th}$ (Sahu et al, 1997a). The conservative errors leave 
room for significant fading. 
However, more data are required to define a clear trend.
The shape of the diffuse emission is difficult to assess, due to higher 
background in the new observations as compared to the previous ones. 
In the following, we concentrate on the V-band images (which show the best signal to noise) to 
obtain a very 
accurate relative astrometry of all point-like objects present in both observations. The 
centroid of each star was computed in the MIDAS environment.
A 
Gaussian function was used to fit the source data, yielding the best positions and associated 
errors. 
Particular care was used in the computation of the centroid of the presumed counterpart of 
GRB970228. 
Although it is not the purpose of this paper to study in detail the diffuse emission discussed 
by Sahu et al.(1997c), close 
attention has been given to its possible influence on the GRB counterpart centering.  A 
blow-up of the region in the March frame, where such a nebulosity is more apparent, is shown in 
figure 2a where green identifies backgroung pixels, averaging 8 counts, while yellow covers the 
range 8.1-8.4 counts (read noise  $\sim 0.75$ and gain $\sim 7.12$).
For comparison the peak source intensity is 14 counts/pixel. The contribution of the diffuse 
emission to the source counts is indeed negligible.
The faintness of the diffuse emission stems clearly from the two orthogonal tracings of Figures 
2b(row) and 2c (column). 
%tridimensional plot of the 20 $\times$ 20 pixel region around the GRB source. 
The shape of the point-source is clearly 
unaffected by the very faint nebulosity.  Therefore, we are confident that the GRB centering is 
not polluted by it.
\\Table 1  lists magnitudes and Gaussian widths measured in the 
March and April observations for all our stars. While objects 2,3,4 and the GRB 970228 
counterpart are 
point-like with very similar width of the Gaussian profiles, object 1 is slightly extended. 
However, we are confident that its centering can be accurately 
performed. We note that the width values, measured for the point-like sources, in April are 
larger than the 
March ones. This must be due to some systematic effect. However, since all the sources are 
affected in 
the same way, such a broadening does not hamper our astrometric comparison.
In the presence of such a systematic difference between the star profiles in the two 
observations, the use of a variable Gaussian profile for accurate star positioning is certainly 
superior to the standand PSF fitting.
The centering errors of the reference stars are between 0.02 and 0.07 pixels, while for the 
candidate 
counterpart of GRB 970228 the uncertainties are $\sim$ 0.07 pixels.   The values of all the 
centroids were corrected 
for the instrument geometrical distorsion, following the detailed procedure outlined by the 
WFPC2 team (Holtzmann et al, 1995). 
\begin{table}
\begin{tabular} {|l|l|c|c|} \hline
{\em Obj.} & {\em mag} &  {\em $FWHM_{x}$} &{\em $FWHM_{y}$}    \\ \hline
1   & $24.3 \pm 0.1$  & 6.6 [6.2] & 5.0 [5.2]  \\ \hline
2   & $22.9 \pm 0.05$ & 3.2 [4.7] & 3.2 [4.1] \\ \hline
3   & $25.8 \pm 0.2$  & 3.0 [4.4] & 2.8 [4.6] \\ \hline
4   & $23.8 \pm 0.1$  & 3.1 [4.3] & 3.1 [4.3] \\ \hline
GRB & $25.7-26.1 \pm 0.2 $ & 3.5 [4.6] & 3.1 [4.2] \\ \hline
\end{tabular}
\caption{ {\it The table lists the magnitudes of the stars labelled in
Fig.1. For the GRB counterpart the two values correspond to the March
and April observations, respectively. In columns 3 and 4 the FWHM components of the gaussian 
functions best
fitting the stars' profiles are given (in pixels, 1pix=0.0455 arcsec). Numbers in
brackets refer to the April observation.}}
\label{}
\end{table}
\begin{figure} 
\centerline{\hbox{
{\bf (a)}
\psfig{figure=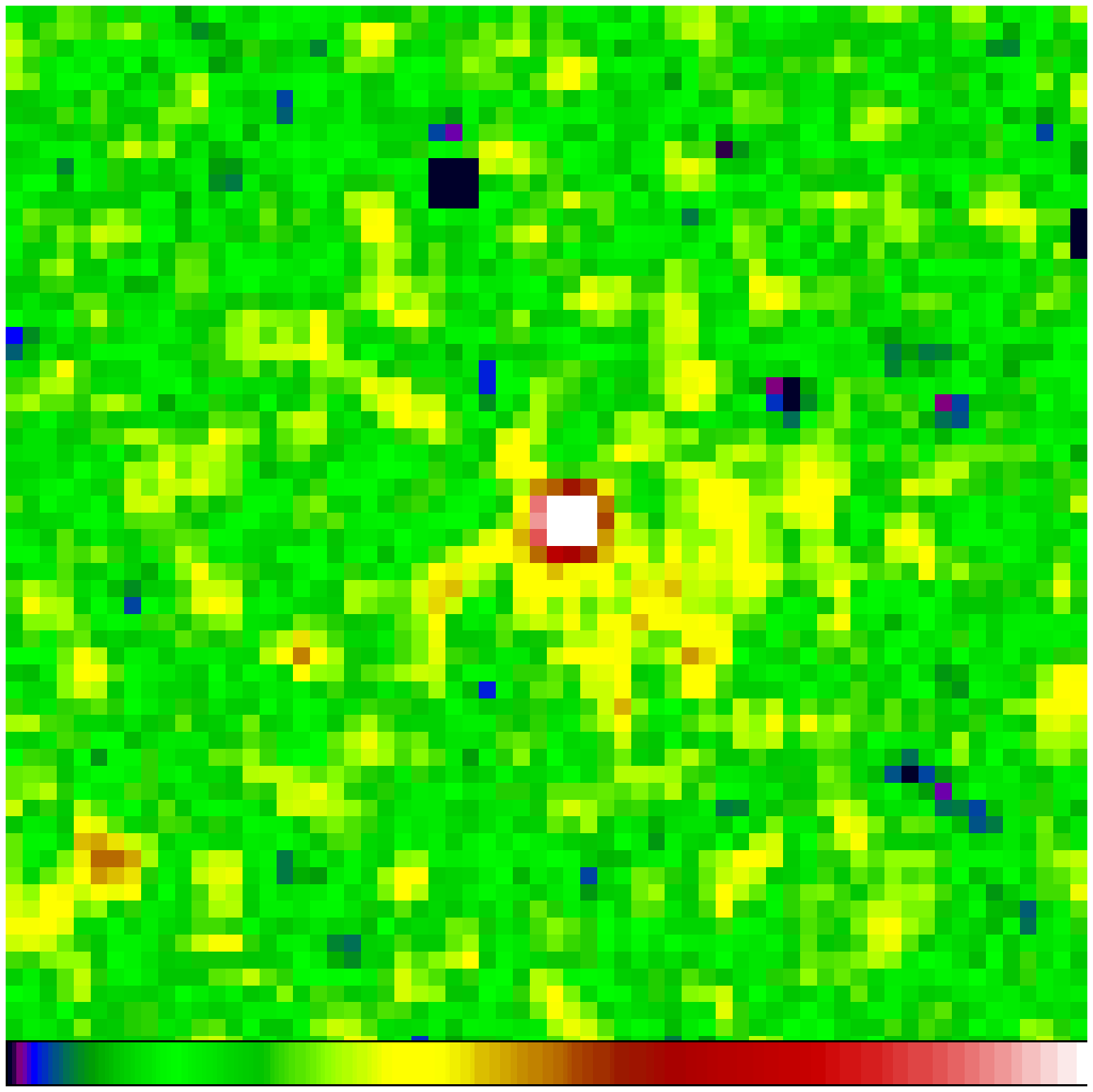,height=8cm,clip=} 
}}
\vspace{0.4cm}
\centerline{\hbox{
{\bf (b)}
\psfig{figure=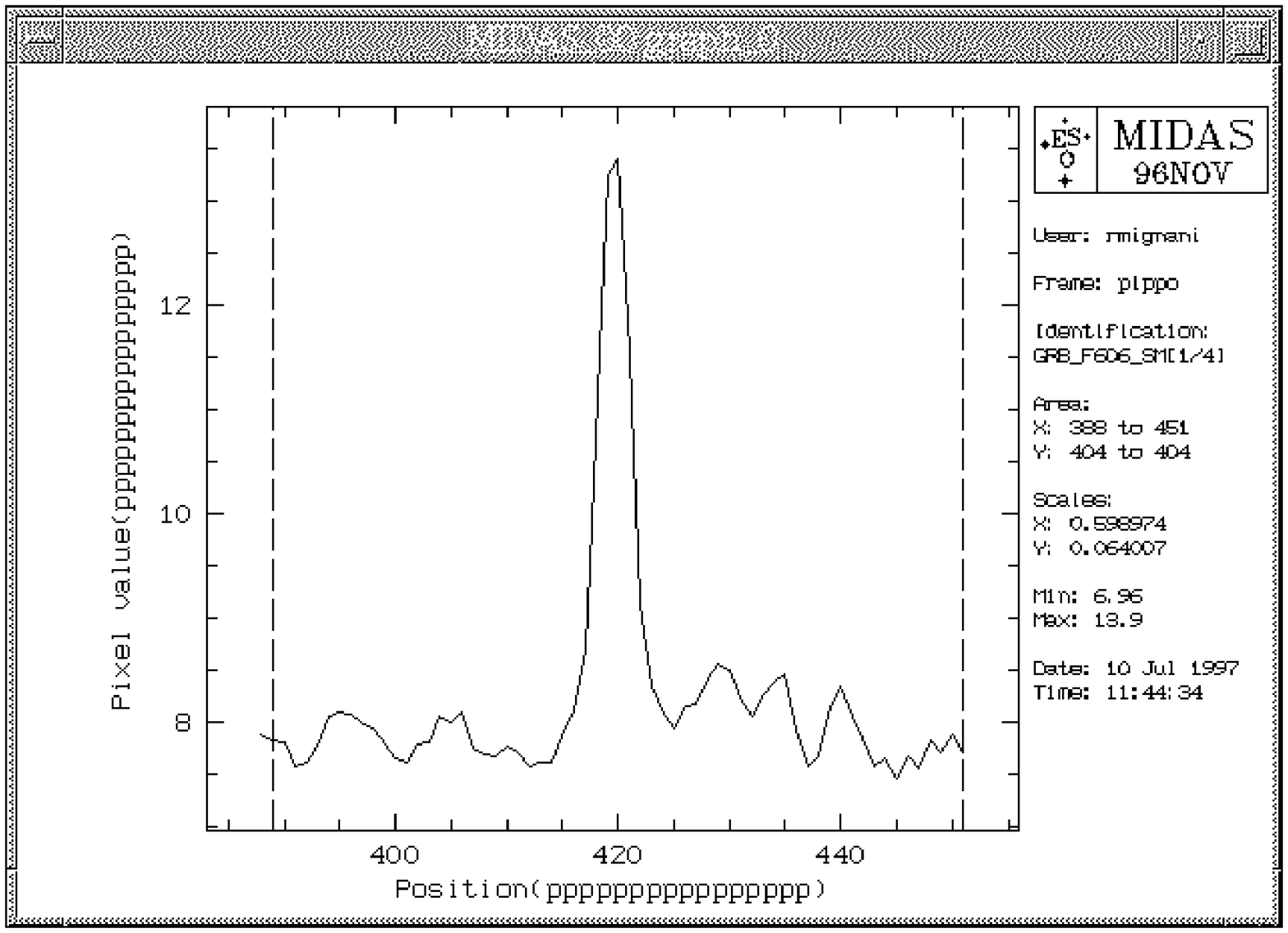,height=3.5cm,width=8cm,clip=}
}}
\vspace{0.2cm}
\centerline{\hbox{
{\bf (c)}
\psfig{figure=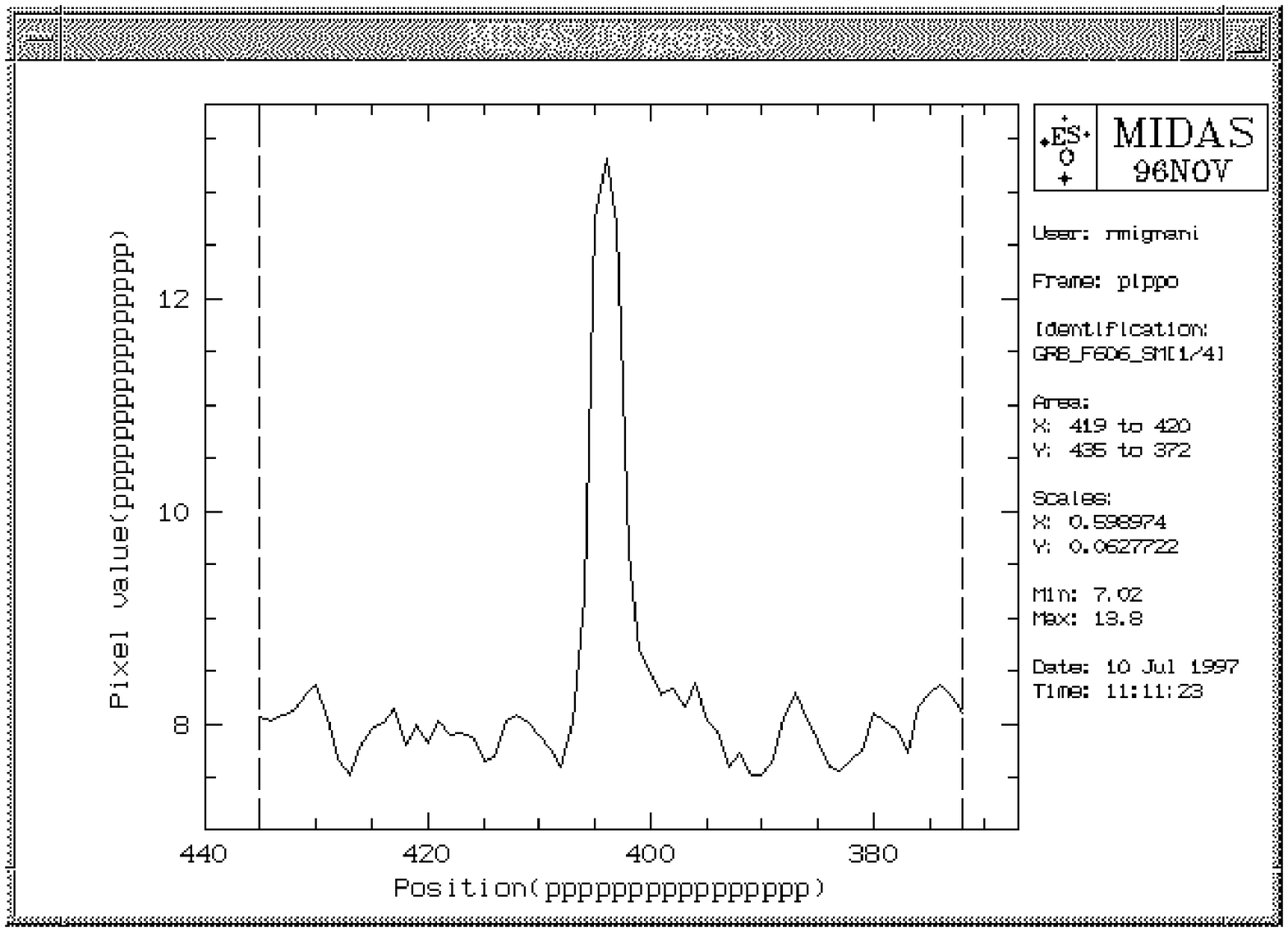,height=3.5cm,width=8cm,clip=}
}}
\caption{(a) Blow-up of the F606W March $26^{th}$ image centered on the GRB 970228 counterpart 
(pixel size is 0.0455 
arcsecs). The square shape of the point source is an artifact due to the color scale used to 
display the very 
faint nebulosity (b) Tracing of the above image along the row passing through the GRB maximum. 
X-axis: each tic corresponds to 5 PC pixels, Y-axis: count/pixel (c) idem for column of maximum} 
 \label{} 
 \end{figure}

The two observations cannot be immediately superimposed because they have been obtained with 
slightly different roll angles and  pointing directions. 
Thus, precise alignment of the two images is needed. This has been performed following the 
procedure used 
by Caraveo et al (1996) to measure the parallactic displacements of Geminga.
%Since the mapping of the geometrical distorsion of the Planetary Camera is very 
%accurate (Holtzmann et al, 1995), the standard 
%PC value of 0.0455 arcsec/pixel can be used as plate scale for all corrected %frames. 
To account for image rotation, we have rotated the source coordinates
according to the known telescope roll angles, ending up with the $X$-axis aligned with 
increasing right 
ascension and the $Y$-axis aligned with increasing declination.   Thus, the statistical weight 
of the 
reference sources was used in establishing the translation factors. Since neither the net shifts 
nor their errors change when object \#1 is excluded, all reference objects were used.
To evaluate globally the errors introduced by the 
procedure described above, we compare the coordinates of our stars in the two images. In the 
absence of significant proper motions, they should be the same, within errors, i.e., their 
residuals should 
cluster around zero. 
 
\section{Results.}

%Table 2 summarizes the results of the centroid cross-comparisons for the $X$ 
%and $Y$ coordinates. The first 
%column contains the field star identification, following the labelling used in 
%Fig 1.
%For each line, $\Delta~X$ ($\Delta~Y$), in pixels, represents the displacement 
%of the star centroid (together with its 
%error)  along the $X (Y)$ axis aligned in right ascension (declination), using 
%the March image as a reference.  
%While the positioning errors originate from the original star centering 
%uncertainties (plus translation 
%accuracy), the actual values of the displacements provide a measure of the 
%precision achieved in the 
%image superposition. 
The results of the centroid cross-comparisons are shown in Figure 3 where, for each star, we 
plotted the April position relative to 
the March one, taken as zero point.
While the positioning errors originate from the original star centering uncertainties (plus 
translation 
accuracy), the actual values of the displacements provide a measure of the precision achieved in 
the 
image superposition. \\Pixel displacements of stars 1,2,3,4 are:  
\\$\Delta \alpha = -0.08 \pm .09, 0.03 \pm .08, 0.07 \pm .10, -0.02 \pm .04$, 
\\$\Delta \delta = -0.05  \pm .09, 0.02 \pm .08 , 0.05 \pm .10,-0.01 \pm .04$,
\\yielding {\it total} April vs. March residuals $< 0.1$ pixels. This gives confidence in the 
overall correctness and 
accuracy of our procedure and provides a global estimate of the final uncertainty. 
The behaviour of the candidate counterpart of GRB 970228 is different: it shows an angular 
displacement of $\Delta ~ \alpha =0.30 \pm 0.1$ pixels and $\Delta ~\delta = -0.26 \pm 
0.1$ pixels,  for a total displacement of $0.40 \pm 0.10$ pixels equivalent to $18 \pm 4.5$ mas 
to the South-East. We 
note that star \#3, of $m_{606}=25.8$, does not show any displacement to within a global error 
budget of 0.1 
pixel.

\begin{figure} 
\centerline{\hbox{
\psfig{figure=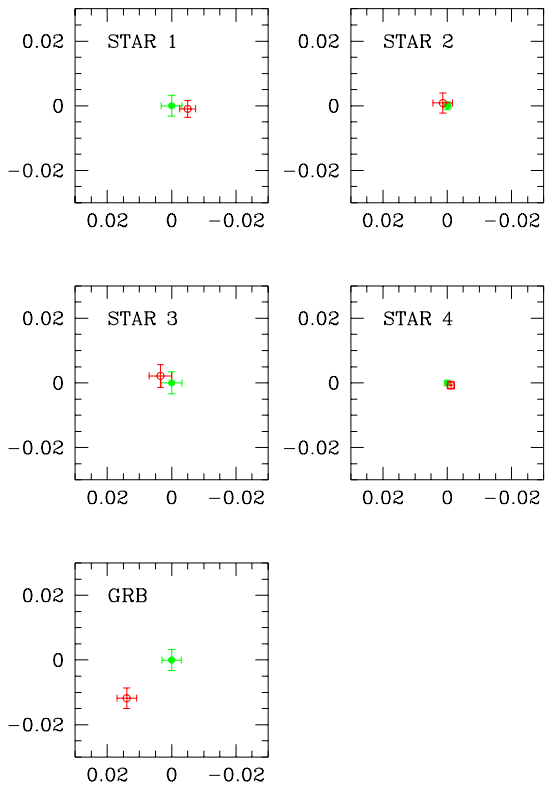,height=11.5cm,clip=}
}}
%\picplace{11.5cm}
\caption{April $7^{th}$ vs March $26^{th}$, 1997 displacements of reference star centroids and 
of the GRB 970228 proposed counterpart.
The March $26^{th}$ positions (identified with filled circles) have been used as zero point. 
Open circles identify the 
positions measured on April $7^{th}$. While the reference stars show no angular displacements, 
the proposed 
counterpart of GRB does.
North to the top, East to the left. Axis units are in arcsec. } 
 \label{} 
 \end{figure}

\section{Discussion.}
In summary, although in need of confirmation, our comparison of the two available HST images 
shows 
evidence that the point-like counterpart of GRB 970228 is moving. \\
Taking the result of our method at face value, the angular displacement found for the optical 
counterpart 
of GRB 970228 is $\sim$ 550 ($\pm$ 140) mas/yr for a constant speed, and implies a transverse 
velocity-distance 
relation of the type : $v (km/sec)=2.7 \times d (pc)$. 
For $v=c$, a value of d $\sim$ 100 kpc is obtained. In this case, the optical emission seen by 
Hubble 36 days after the 
burst  would be due to a relativistic plasmoid ejected from a compact source in a jet-like 
geometry. 
A transverse velocity of few hundreds km/sec, typical of radio pulsars (Lyne \& Lorimer, 1994),  
would correspond to a distance of 
$\sim$ 100 pc.  \\ 
The WFPC2 has already measured comparable V-band fluxes from nearby neutron stars 
such as Geminga (Bignami et al, 1996) and  PSR0656+14 (Mignani et al, 1997), two middle-aged 
isolated neutron stars (INSs)
with faint optical emission. 
However, we note that the colors of the GRB candidate may not follow a simple Rayleigh-Jeans 
extrapolation of the 
spectrum, also owing to possible circumstellar nebulosity. Moreover, different emission 
mechanisms 
(see Tavani, 1997) should also be considered, possibly in the context of older and cooler 
isolated neutron stars, 
emitting optical radiation because of a burst-driven heat release or residual accretion. \\
Constraints on a galactic population of compact objects  producing GRBs can be deduced from the 
isotropic and 
apparently non-homogeneous  distribution of GRBs obtained by BATSE (e.g. Fishman \& Meegan, 
1991). Arguments against a 
local disk population of GRBs (Paczynski, 1991; Mao \& Paczynski, 1992) are mostly based on the 
difficulty of reconciling the observed GRB 
isotropic and non-homogeneous distribution (Meegan et al, 1992; Briggs et al, 1996) 
with compact objects distributed as Population I stars.  
However, a recent  analysis of BATSE data indicates that non-homogeneity  may not apply to the 
whole 
sample of GRBs (e.g. Kouveliotou et al, 1996). Several  classes  of GRB sources with different 
spatial distributions may exist. 
A fast moving counterpart of 
GRB 970228 can be ascribed to  a local population  of compact  objects with an  isotropic and 
homogeneous  
spatial distribution. The relatively long duration of GRB 970228 ($\sim$ 80 s) and its likely 
low value of the 
average spectral hardness (Costa et al, 1997) are consistent with being a burst belonging to the 
homogeneous population 
detected by BATSE (e.g. Kouveliotou et al, 1996). 
Alternatively, a relativistic plasmoid at d $\sim$ 100 kpc may be consistent with observations. 
Relatively rapid 
fading within a time scale of 1-2 months after the burst is expected for a relativistic 
plasmoid. A slower 
decay of emission from the point-like object may result from residual surface or disc emission 
of a 
compact object.  More interpretative work is needed to clarify these issues. \\
In any case, any proper motion of the GRB 970228 counterpart would prove its galactic nature. 
The 
reported fading of the surrounding optical nebulosity seems to support a galactic origin of the 
source. However, this fading has now been put in question (Fox et al, 1997).  GRB 970228 
may be representative of a large fraction, if not all, of GRB sources. 
In view of the importance of the topic and of the preliminary nature of our result, which 
stretches HST capabilities to their limit, further observations are needed. They will both 
extend the time span for 
the proper motion measurement and gauge the source  luminosity evolution. More X-ray afterglows 
from other GRBs may be detected in the near future by BeppoSAX, and rapid optical follow-up 
observations will further constrain the nature of GRB sources.

%\vspace{1.0cm}
%
%\noindent
%{\it Acknowledgments.} \\
%We acknowledge the NASA/STScI policy for rapid archiving of HST data which made 
% possible our study.

\end{document}